\documentclass[dvips,12pt]{article}
\usepackage{graphicx}
\usepackage{amssymb}
\usepackage{graphicx}
\usepackage{dcolumn}
\usepackage{amsmath}
\usepackage{lscape}
\usepackage{longtable}

\begin{document}
\textwidth 16cm
\newcommand{\bd}{\begin{document}}
\newcommand{\ed}{\end{document}}
\newcommand{\bc}{\begin{center}}
\newcommand{\ec}{\end{center}}
\newcommand{\bfr}{\begin{flushright}}
\newcommand{\efr}{\end{flushright}}
\newcommand{\lt}{\left}
\newcommand{\rt}{\right}
\newcommand{\vs}{\vspace}
\newcommand{\hs}{\hspace}
\newcommand{\beq}{\begin{equation}}
\newcommand{\eeq}{\end{equation}}
\newcommand{\lb}{\linebreak}
\newcommand{\pb}{\pagebreak}
\newcommand{\mb}{\makebox}
\newcommand{\fb}{\framebox}
\newcommand{\mc}{\multicolumn}
\newcommand{\ben}{\begin{enumerate}}
\newcommand{\een}{\end{enumerate}}
\newcommand{\bit}{\begin{itemize}}
\newcommand{\eit}{\end{itemize}}
\newcommand{\ol}{\overline}
\newcommand{\un}{\underline}
\newcommand{\lefq}{\lefteqn}
\newcommand{\ba}{\begin{array}}
\newcommand{\ea}{\end{array}}
\newcommand{\beqa}{\begin{eqnarray}}
\newcommand{\eeqa}{\end{eqnarray}}
\newcommand{\beqas}{\begin{eqnarray*}}
\newcommand{\eeqas}{\end{eqnarray*}}
\newcommand{\bfg}{\begin{figure}}
\newcommand{\efg}{\end{figure}}
\newcommand{\bds}{\begin{displaymath}}
\newcommand{\eds}{\end{displaymath}}
\newcommand{\btb}{\begin{tabbing}}
\newcommand{\etb}{\end{tabbing}}
\bc {\huge $su(1,1)\simeq so(2,1)$ Lie Algebraic Extensions of the Mie-type Interactions with Positive Constant Curvature } \ec

\vs{1cm}

\bc
{\it \"Ozlem Ye\c{s}ilta\c{s}{\footnote {e-mail : yesiltas@gazi.edu.tr}\\
Department of Physics, Faculty of Science,
Gazi University,
06500 Ankara, Turkey\\
\vspace{.16cm}

}} \ec
\vs{1cm}

\begin{abstract}
The Schr\"{o}dinger equation in three dimensional space with constant positive curvature is studied for the Mie potential. Using analytic polynomial solutions, we have obtained whole spectrum of the corresponding system. With the aid of factorization method, ladder operators are obtained within the variable and function transformations. Using ladder operators, we have given the generators of $so(2,1)$ algebra and the Casimir operator which are related to the Mie Oscillator on the positive curvature.

\end{abstract}

\noindent { keyword:positive curvature, mie potential}  \\


\section{Introduction}	
Recently, quantum mechanics in curved spherical spaces as a fundamental problem has become a subject of intense research efforts \cite{car, sad, sad1, pah, nieo, in, ham}. The notion of the constant curvature and the accidental degeneracy first began with Schr\"{o}dinger \cite{sch}, Infeld \cite{infeld}, Stevenson \cite{ste}. Essential advances of these systems with accidental degeneracy have been made by Nishino \cite{N}, Higgs \cite{H}, Leemon \cite{L}. It has been found that the complete degeneracy of the energy of the Coulomb problem and harmonic oscillator on the three dimensional sphere in the orbital and azimuthal quantum number is caused by an additional integral of motion. At the same time, some papers on curved spherical spaces are concerned with some applications of physics such as linear and non-linear optics \cite{ba}, quantum dots \cite{dot1, dot2}. Furthermore, in \cite{Lop}, the authors studied liquid crystals using spherical geometries. Thus, molecular potentials such as Mie type interactions may be an interesting candidate for the topological applications of some molecules. Symmetry groups have come to play an important role in quantum physics. On the other hand, symmetry algebras enable one to understand the degenerate energy eigen-states of a system, exact solvability of the spectrum of a quantum system usually indicates the presence of symmetry. In \cite{mesa}, symmetry algebras are studied within exact solvability and $so(2,2)$ algebras.

To our knowledge, there has not been studied Mie-type interactions, which are used to determine molecular structures \cite {mol1, mol2, s1, sever, same, ag}, in constant positive curvature. Hence, this study is concerned with the extension of Mie potential to the spherical coordinates with spaces of constant curvature and to the symmetry algebras determining $so(2,1)$ algebra for the Hamiltonian which is factorized and defined as Mie interactions on constant positive curvature. This paper is organized as follows. The Mie potential in  spherical coordinates with spaces of constant curvature is given in Section $2$. Section $3$ presents the solutions of the eigenvalue equation which is derived from the Schr\"{o}dinger equation with Laplace-Beltrami operator. Section $4$ is assigned to discuss symmetry algebras which are more general than the potential algebras for the corresponding system.
\section{Mie Potential on the Constant Curvature}
We will attend to the case of the three dimensional space of constant positive curvature which is geometrically given on the three dimensional sphere of radius $R$, $\mathbb{S}^{3}$ embedded into the four dimensional Euclidean space when the equation of $\mathbb{S}^{3}$ has the form
\begin{equation}\label{1}
  \mathbb{S}^{3}=\{(\zeta_{0}, \zeta_{i}) \in \mathbb{R}^{4}: \zeta^{2}_{0}+\zeta_{i}\zeta_{i}=R^{2}\}
\end{equation}
where $i=1, 2, 3$ in the tangent space $x_{i}$ are the coordinates and  $\zeta_{i}$ is
\begin{equation}\label{2}
    \zeta_{i}=\frac{x_{i}}{\sqrt{1+\frac{r^{2}}{R^{2}}}}
\end{equation}
\begin{equation}\label{3}
     \zeta_{0}=\frac{R}{\sqrt{1+\frac{r^{2}}{R^{2}}}}.
\end{equation}
The spherical coordinates are given by
\begin{eqnarray}
  \zeta_{1} &=& R \sin \psi \sin \theta \cos \phi \label{a}\\
  \zeta_{2} &=& R \sin \psi \sin \theta \sin \phi \label{b} \\
  \zeta_{3} &=& R \sin \psi \cos \theta \label{c}\\
  \zeta_{4} &=& R \cos \psi   \label{d}
\end{eqnarray}
where $0\leq \psi \leq \pi$, $0\leq \theta\leq \pi$, $0\leq \phi <2\pi$. Differentiating with respect to the arbitrary angles $\psi, \theta, \phi$ gives a four dimensional vector and the squared length of this vector is
\begin{equation}\label{4}
    ds^{2}=R^{2}\left(d\psi^{2}+\sin^{2} \psi(d\theta^{2}+\sin^{2} \theta d\phi^{2})\right)
\end{equation}
which is called as Robertson-Walker metric for the positive curvature $\kappa=1$. Define $r^{2}=x^{2}_{1}+x^{2}_{2}+x^{2}_{3}$ and the potential $V(r)$ which is known as Mie potential \cite{s1, sever, same} given by,
\begin{equation}\label{5}
    V(r)= \varepsilon \left(\frac{k}{l-k}\left(\frac{a}{r}\right)^{l}-\frac{l}{l-k}\left(\frac{a}{r}\right)^{k}\right); ~~l=2k; ~~k=1,
\end{equation}
where $\varepsilon$ is the interaction energy between the atoms in a molecule, $a$ is the coordinate of the interaction, $l > k$. A special case that is $k=1$  performed as
\begin{equation}\label{6}
    V(r)=V_{0}\left(\frac{1}{2}\left(\frac{a}{r}\right)^{2}-\frac{a}{r}\right), ~~V_{0}=2\varepsilon k.
\end{equation}
Inserting the above dependence of $r$ on $\zeta$ gives
\begin{equation}\label{7}
    V(\zeta)=V_{0}\left(\frac{a^{2}}{2}\frac{1-\frac{\zeta^{2}}{R^{2}}}{\zeta^{2}}-a\frac{\sqrt{1-\frac{\zeta^{2}}{R^{2}}}}{\zeta}\right),
\end{equation}
or we may give $V(\psi)$ as
\begin{equation}\label{70}
    V(\psi)=V_{0}\left(\frac{1}{2}\left(\frac{a}{R \tan \psi}\right)^{2}-\frac{a}{R \tan \psi}\right).
\end{equation}
In fact, this potential (\ref{70}) is known as trigonometric Rosen- Morse I potential \cite{suk}.
\section{Eigenvalue Equation and Solutions}
Here we give  the Schr\"{o}dinger equation for (\ref{7}) on the constant curvature,
\begin{equation}\label{8}
   \left( -\frac{\hbar^{2}}{2\mu}\Delta+V\right) \Psi=E \Psi,
\end{equation}
where $\Delta$ is the Laplace-Beltrami operator which is a restriction of the Laplace operator on the  sphere, then we have the following formula for
\begin{equation}\label{9}
    \Delta=\frac{1}{\sqrt{g}}\sum^{3}_{i,k=1}\frac{\partial}{\partial x^{i}} \left(\sqrt{g}g^{ik}\frac{\partial}{\partial x^{k}}\right),
\end{equation}
and define the metric which is
\begin{equation}\label{10}
    ds^{2}=g_{ik} dx^{i} dx^{k}
\end{equation}
where $g=det|g_{ik}|$ and by the chain rule $ g^{ik}=(g_{ik})^{-1}$. Thus, using (\ref{a}), (\ref{b}), (\ref{c}) and (\ref{d}), (\ref{9}), Schr\"{o}dinger equation takes the form
\begin{equation}\label{11}
\begin{split}
   \left (\frac{1}{\sin^{2} \psi}\frac{\partial}{\partial \psi} \sin^{2} \psi \frac{\partial}{\partial \psi}\right)\Psi +
  \frac{2\mu R^{2}}{\hbar^{2}} \\ \left(E-\frac{\hbar^{2}}{2\mu R^{2}}\frac{m(m+1)}{\sin^{2}\psi}-V_{0}\left(\frac{1}{2}\left(\frac{a}{R \tan \psi}\right)^{2}-\frac{a}{R \tan \psi}\right)\right)\Psi & =0.
  \end{split}
\end{equation}
Using a transformation of the wave-function in (\ref{11})
\begin{equation}\label{12}
    \Psi(\psi)=\frac{\phi(\psi)}{\sin \psi}
\end{equation}
and
\begin{eqnarray}
  C_{1} &=& \frac{2\mu R^{2}}{\hbar^{2}}\left(E+\frac{a^{2}V_{0}}{2R^{2}}\right) \label{e} \\
  C_{2} &=& \frac{2\mu R^{2}}{\hbar^{2}} \left(\frac{\hbar^{2}m(m+1)}{2\mu R^{2}}+\frac{V_{0}a^{2}}{2R^{2}}\right) \label{f} \\
  C_{3} &=& \frac{2\mu R^{2}}{\hbar^{2}} \frac{aV_{0}}{R} \label{g}
\end{eqnarray}
(\ref{11}) turns into
\begin{equation}\label{13}
    \phi^{''}+(C_{1}+2+C_{3}\cot \psi-(C_{2}+2)\csc^{2}\psi)\phi=0.
\end{equation}
Another transformation of the variables which are
\begin{equation}\label{14}
    \phi(\psi)=e^{-\alpha \psi/2} F(\psi), ~~~~z=\cot \psi
\end{equation}
lead to
\begin{equation}\label{15}
\begin{split}
    (1+z^{2})^{2}F^{''}(z)+2(1+z^{2})(\alpha+z)F^{'}(z)&+\\ \left(C_{3}z-(C_{2}+2)(1+z^{2})+C_{1}+2+\frac{\alpha^{2}+1}{4}\right)F(z)=0.
\end{split}
\end{equation}
Finally, we shall use an ansatze in above equation as
\begin{equation}\label{16}
    F(z)=(1+z^{2})^{-\frac{1-\beta}{2}}f(z),
\end{equation}
then we can obtain
\begin{equation}\label{17}
\begin{split}
    f^{''}(z)+\frac{\alpha+2\beta z}{1+z^{2}}f^{'}(z)+\frac{1}{(1+z^{2})^{2}}(C_{1}-C_{2}+\beta+\frac{\alpha^{2}-3}{4}&+\\
    \left(C_{3}-\alpha+\alpha \beta)z+ (\beta^{2}-\beta-C_{2}-2)\right)f(z) =0.
    \end{split}
\end{equation}
Let us arrange the coefficient of $f(z)$ in (\ref{17}) as
\begin{equation}\label{18}
    \frac{1}{1+z^{2}}\left(C_{1}+\frac{\alpha^{2}+5}{4}+2\beta-\beta^{2}+z(C_{3}-\alpha+\alpha \beta)\right)-2-C_{2}-\beta+\beta^{2}
\end{equation}
and the coefficients of $z$ and $z^{2}$ can be terminated in (\ref{18})  if
\begin{eqnarray}
  C_{3}-\alpha(1-\beta) &=& 0 \label{c3} \\
  C_{1}+2\beta-\beta^{2}+\frac{\alpha^{2}+5}{4} &=& 0 \label{c1}.
\end{eqnarray}
Then, we may continue to search to obtain a hypergeometric equation, hence we use
\begin{equation}\label{19}
    z=i t,~~~~\alpha\rightarrow i\alpha
\end{equation}
in (\ref{17}) and this yields
\begin{equation}\label{20}
    (1-t^{2})f^{''}(t)-(\alpha+2\beta t)f^{'}(t)+(\beta(1-\beta)+2+C_{2})f(t)=0.
\end{equation}
Jacobi differential equation is given as \cite{abra}
\begin{equation}\label{21}
    (1-x^{2})y^{''}(x)+(\textbf{b}-\textbf{a}-(\textbf{a}+\textbf{b}+2)x)y^{'}(x)+n(n+\textbf{a}+\textbf{b}+1)y(x)=0.
\end{equation}
We now compare (\ref{20}) and (\ref{21}) in order to express the solutions $f(t)$ in terms of Jacobi polynomials, and then we have
\begin{equation}\label{22}
    \textbf{a}=\frac{2\beta-\alpha-2}{2}, ~~~~\textbf{b}=\frac{2\beta+\alpha-2}{2}.
\end{equation}
Thus, our solutions $f(t)$ can be given as $f(t)=P_{n}^{(\mathbf{a},\mathbf{b})}(t)$. Moreover, let us substitute
\begin{equation}\label{23}
    C_{2}+2=j(j+1)
\end{equation}
in (\ref{17}), one can see that
\begin{equation}\label{24}
    n(n+2\beta-1)=\beta(1-\beta)+j(j+1).
\end{equation}
Shifting $n$ to $n\rightarrow n-1$ and using (\ref{c3}); we find the followings which are $n$-dependent constants:
\begin{equation}\label{25}
    \alpha_{n}=\frac{C_{3}}{n+j}
\end{equation}
\begin{equation}\label{26}
    \beta_{n}=1-(n+j).
\end{equation}
Finally, (\ref{c1}) leads to find our energy eigenvalues as
\begin{equation}\label{27}
    E_{n}=\frac{\hbar^{2}}{2\mu R^{2}}\left((n+j)^{2}-\frac{C^{2}_{3}}{4(n+j)^{2}}-\frac{\mu a^{2}V_{0}}{\hbar^{2}}-\frac{9}{4}\right).
\end{equation}
We may also give $j$ in terms of parameters of the potential as,
\begin{equation}\label{28}
    j=-\frac{1}{2}\pm \sqrt{m(m+1)+\frac{\mu  V_{0}a^{2}}{\hbar^{2}}+\frac{7}{4}}.
\end{equation}
And, we can write the un-normalized eigenfunction solutions of (\ref{8}) which are complex as
\begin{equation}\label{29}
    \Psi(\psi)=\frac{N}{\sin \psi} e^{-i\alpha \psi/2}(1+\cot^{2} \psi)^{-\frac{1-\beta}{2}} P^{(\textbf{a},\textbf{b})}_{n}(-i\cot \psi).
\end{equation}
where $P^{(\textbf{a},\textbf{b})}_{n}$ are corresponding Jacobi polynomials. On the other hand, in the limit of $R\rightarrow\infty$,
\begin{equation}\label{30}
    R\rightarrow\infty, ~~~~E_{n}\rightarrow -\frac{\mu^{2}a^{2}V^{2}_{0}}{\hbar^{4}}\frac{1}{(n+j)^{2}}
\end{equation}
which means we have energy eigenvalues (\ref{30}) in flat space and this agrees with the results in \cite{sever}. When $\psi\rightarrow 0$, $\psi\rightarrow \pi$, we obtain $\Psi_{n}\rightarrow 0$.

\section{Factorization and Algebra}
Let us re-consider (\ref{13}) which is
\begin{equation}\label{31}
    \frac{d^{2}\phi}{d\psi^{2}}+\left(\epsilon-\frac{j(j+1)}{\sin^{2} \psi}+2A\cot \psi \right)\phi=0,
\end{equation}
where we used $C_{1}+2=\epsilon$ and $A=-\frac{\mu R aV_{0}}{\hbar^{2}}$. If we perform the changes of variable and also of  function,
\begin{equation}\label{32}
    \tan\frac{\psi}{2}=e^{-i \frac{y-\pi}{2}}, ~~~~\phi(z)=\frac{1}{\sqrt{\cosh z}} \chi
\end{equation}
in (\ref{31}), we get
\begin{equation}\label{33}
    \chi^{''}(y)-\frac{\epsilon-\frac{1}{4}+2iA\cos y}{\sin^{2} y}\chi(y)+(j+\frac{1}{2})\chi(y)=0,
\end{equation}
which is known as type A operators in the book by Miller \cite{miller}. In \cite{miller}, type A factorization tells us about a linear second-order differential equation like (\ref{31}) can be factorized if (\ref{31}) is written as
\begin{equation}\label{34}
   \mathcal{ A}^{+}(j+1)\mathcal{A}^{-}(j+1)Y(\epsilon, j)=(\epsilon-R(j+1))Y(\epsilon, j)
\end{equation}
\begin{equation}\label{35}
    \mathcal{A}^{-}(j)\mathcal{A}^{+}(j) Y(\epsilon, j)=(\epsilon-R(j)) Y(\epsilon, j)
\end{equation}
where
\begin{equation}\label{36}
    \mathcal{A}^{\pm}=\pm \frac{d}{dy}+p(y,j).
\end{equation}
Here, $\mathcal{A}^{\pm}$ are known as ladder operators which read
\begin{equation}\label{37}
    Y(\epsilon, j\pm 1)=\mathcal{A}^{\mp} Y(\epsilon, j)
\end{equation}
and satisfy $(\mathcal{A}^{+}y_{1}, y_{2})= (y_{1}, \mathcal{A}^{-}y_{2})$. Let  $p(y, j)$ be
\begin{equation}\label{38}
    p(y, j)=(j+s)\cot y+\frac{t}{\sin y}.
\end{equation}
Then, plugging (\ref{38}) into (\ref{35}) and (\ref{36}), we have
\begin{equation}\label{39}
    \mathcal{A}^{-}\mathcal{A}^{+}Y=-Y^{''}+\frac{(j+s)(j+s+1)+t^{2}+2t(j+s+1/2)\cos y}{\sin^{2} y}Y+\epsilon Y=0
\end{equation}
where we use $s, t$ are real constants. And we may give the ladder operators and $R(j)$ as
\begin{equation}\label{40}
   \mathcal{ A}^{\pm}=\pm \frac{d}{dy}+(j+s)\cot y+\frac{t}{\sin y}
\end{equation}
\begin{equation}\label{41}
    R(j)=(j+s)^{2}.
\end{equation}
Square integrability of the solutions requires $j=0,1,2,...\ell$, $\epsilon=R(\ell)$. To define the unknown parameters $s$ and $t$, we can compare (\ref{39}) and (\ref{33}) as follows:
\begin{equation}\label{42}
    t^{2}+(j+s)(j+s+1)=\epsilon^{2}-\frac{1}{4}
\end{equation}
\begin{equation}\label{43}
    t(j+s+1/2)=iA.
\end{equation}
Using (\ref{43}) and substituting it into (\ref{42}), we obtain
\begin{equation}\label{44}
    \epsilon=-A^{2}/t^{2}+t^{2}.
\end{equation}
If we consider (\ref{31}), $\epsilon$ is given by \cite{miller},
\begin{equation}\label{45}
    \epsilon=R(j+1)=(j+1)^{2}-\frac{A^{2}}{(j+1)^{2}}.
\end{equation}
Hence, we have two $\epsilon$ expressions, if we equate (\ref{44}) and (\ref{45}), we get
\begin{equation}\label{46}
    t=\pm (\ell+1), ~~~~j+s+1/2=\pm i\frac{A}{\ell+1}
\end{equation}
\begin{equation}\label{47}
    t=\pm i \frac{A}{\ell+1},~~~~j+s+1/2=\pm (\ell+1).
\end{equation}
Now, we may consider the Lie algebra generators which we call $X_{1}$, $X_{2}$, $\mathcal{O}_{1}$, $\mathcal{O}_{2}$. The new variables $y_{1}, y_{2}$ can be used in the eigenfunction which is \cite{mesa}
\begin{equation}\label{48}
    \Phi(\psi, y_{1},y_{2})\sim e^{i \nu y_{1}} \phi(\psi) e^{i \kappa y_{2}}
\end{equation}
where
\begin{equation}\label{49}
    \nu=\pm \frac{A}{\ell+1}, ~~~~\kappa= \pm (\ell+1).
\end{equation}
The operators $X_{1}, X_{2}$ act on the eigenfunctions as given below
\begin{equation}\label{51}
    X_{1}\Phi_{\nu, \kappa}=\nu \Phi_{\nu, \kappa}, ~~~~X_{2}\Phi_{\nu, \kappa}=\kappa \Phi_{\nu, \kappa}
\end{equation}
where
\begin{equation}\label{52}
    X_{1}=-i \frac{\partial}{\partial y_{1}},~~~~X_{2}=-i \frac{\partial}{\partial y_{2}}.
\end{equation}
One can map each ladder operator into the new one by taking (\ref{32}) into the consideration as
\begin{equation}\label{53}
    \bar{\mathcal{A}}^{\pm}=\mp i \sin \psi \frac{d}{d\psi}+i (j+s\pm 1/2)\cos \psi+t \sin \psi
\end{equation}
which lead to get generators $\mathcal{O}_{1}$, $\mathcal{O}_{2}$ as
\begin{equation}\label{54}
   \mathcal{ O}^{\pm}_{1}=e^{\pm i y_{1}}\left(\pm \sin \psi \frac{\partial}{\partial \psi}-i \cos \psi \frac{\partial}{\partial y_{1}}-\sin \psi \frac{\partial}{\partial y_{2}}\right)
\end{equation}
\begin{equation}\label{55}
   \mathcal{ O}^{\pm}_{2}=e^{\pm i y_{2}}\left(\pm \sin \psi \frac{\partial}{\partial \psi}-i \cos \psi \frac{\partial}{\partial y_{2}}-\sin \psi \frac{\partial}{\partial y_{1}}\right)
\end{equation}
where we use
\begin{eqnarray}
  \mathcal{O}^{\pm}_{1} &=& -i e^{\pm i y_{1}} \mathcal{\bar{A}}_{1}^{\mp}, ~~~~t=\pm (\ell+1), ~~s+t+1/2=\pm i \frac{A}{\ell+1} \\
  \mathcal{O}^{\pm}_{2} &=& -i e^{\pm i y_{2}} \mathcal{\bar{A}}_{2}^{\mp},~~~~t=\pm i\frac{A}{\ell+1},~~j+s+1/2=\pm (\ell+1).
\end{eqnarray}
One can look at the commutation relations satisfied by the generators above
\begin{equation}\label{56}
    [X_{1}, \mathcal{O}_{\pm}]=\pm \mathcal{O}_{\pm},~~~~[\mathcal{O}_{+}, \mathcal{O}_{-}]=-2X_{1}
\end{equation}
and Casimir operators whose action on the $\Phi_{\nu, \kappa}$ is given by
\begin{equation}\label{57}
    \mathcal{C}\Phi_{\nu, \kappa}=j(j+1) \Phi_{\nu, \kappa}
\end{equation}
equal to
\begin{equation}\label{58}
    \mathcal{C}=-\mathcal{O}^{+}_{1}\mathcal{O}^{-}_{1}+X_{1}(X_{1}-1)=-\mathcal{O}^{+}_{2}\mathcal{O}^{-}_{2}+X_{2}(X_{2}-1).
\end{equation}

\section{Conclusion}
We have studied the Mie potential in spherical curved spaces with constant positive curvature through  both analytical and algebraic approaches including the Infeld factorization. It is seen that, Mie potential is transformed into a Rosen- Morse I-like potential in spherical spaces. We have obtained spectrum and eigenfunctions of the system using polynomial solutions. Our results agree with \cite{sever} if the limit $R\rightarrow \infty$ is used for the solutions. Then, we have used factorization method  to determine the algebra for the system. We remind that the conserved quantity, eigenvalue of the Casimir operator is the potential parameter. Using ladder operators which we obtained with factorization method, we have constructed the $so(2,1)$ algebra for the Mie potential in spherical spaces. In the study of \cite{mar}, $N$ dimensional Mie potential is studied and $su(1,1)$ algebra generators were represented by infinite-dimensional Hilbert subspaces of the radial quantum states. Hence we have extended the system to $so(2,1)$ algebra through a more general procedure. We give graphs of the potentials $V(r)$ and $V(\psi)$ for some specific molecules $CH, NO, N_{2}$ and the data for the potential parameters are given in \cite{M}. Figure $1$ shows that the Mie potential which is in flat space, a standard Coulombic potential energy plus electronic kinetic energy for each molecule, and figure $2$ shows that the Mie potential in spherical spaces with constant positive curvature which may be interesting that the potential takes the form of well with two minima.

Finally, relativistic equations and factorization procedure in spherical spaces within the construction of the algebra may be studied in the future.

\begin{figure}[!htb]
\centering
\includegraphics[scale=.7]{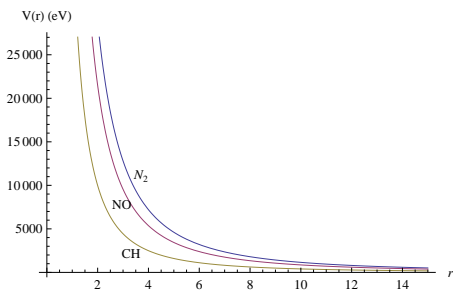}
\caption{Graph of $V(r)$ in (\ref{6}).}
\label{fig:digraph}
\end{figure}

\begin{figure}[!htb]
\centering
\includegraphics[scale=.7]{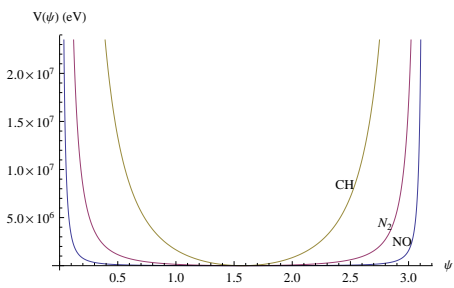}
\caption{Graph of $V(\psi)$ in (\ref{70}).}
\label{fig:digraph}
\end{figure}


\end{document}